\documentclass[aps,reprint,twocolumn,superscriptaddress,10pt]{revtex4-2}
\usepackage{siunitx}
\usepackage{amsfonts}
\usepackage{amssymb}
\usepackage{bm}
\usepackage{graphicx}
\usepackage{dcolumn}
\usepackage{mciteplus}
\usepackage{euscript}
\usepackage{multirow}
\usepackage{color}
\usepackage{textcomp}
\usepackage{gensymb}
\usepackage{float}
\usepackage{enumitem}
\usepackage{xcolor}
\usepackage{mhchem}

\graphicspath{{./}{../figs/}{../psd/}}

\newcommand{\rfig}[1]{Fig.~\ref{#1}}
\newcommand{\rFig}[1]{Figure~\ref{#1}}
\newcommand{\rtbl}[1]{Table~\ref{#1}}

\newcommand{\req}[1]{Eq.~(\ref{#1})}

\newcommand{\ef}{E_{\rm F}}
\newcommand{\ueff}{U}
\newcommand\abinitio{\emph{ab initio}}
\newcommand\etal{\textit{et al.}}
\newcommand{\mub}{\mu_\text{B}}
\newcommand{\tc}{T_\text{C}}
\newcommand{\vvi}{\text{VI}_3}
\newcommand{\cri}{\text{CrI}_3}
\newcommand{\cgt}{\text{CrGeTe}_3}
\newcommand{\tj}{\tilde{J}}
\newcommand{\bfq}{{\bf q}}

\begin{document}

\title{Magnetic interactions and spin excitations in van der Waals ferromagnet $\vvi$}
\author{Elijah Gordon}
\affiliation{Ames Laboratory, U.S.~Department of Energy, Ames, Iowa 50011}
\author{V. V. Mkhitaryan}
\affiliation{Ames Laboratory, U.S.~Department of Energy, Ames, Iowa 50011}
\author{Haijun Zhao}
\affiliation{Ames Laboratory, U.S.~Department of Energy, Ames, Iowa 50011}
\affiliation{School of Physics, Southeast University, Nanjing 211189, China.}
\author{Y. Lee}
\affiliation{Ames Laboratory, U.S.~Department of Energy, Ames, Iowa 50011}
\author{Liqin Ke}
\affiliation{Ames Laboratory, U.S.~Department of Energy, Ames, Iowa 50011}

\begin{abstract}
 Using a combination of density functional theory (DFT) and spin-wave theory methods, we investigate the magnetic interactions and spin excitations in semiconducting VI$_3$.
 Exchange parameters  of monolayer, bilayer, and bulk forms are evaluated by mapping the magnetic energies of various spin configurations, calculated using DFT+$U$, onto the Heisenberg model.
 The intralayer couplings remain largely unchanged in three forms of VI$_3$, while the interlayer couplings show stronger dependence on the dimensionality of the materials.
 We calculate the spin-wave spectra within a linear spin-wave theory and discuss how various exchange parameters affect the magnon bands.
 The magnon-magnon interaction is further incorporated, and the Curie temperature is estimated using a self-consistently renormalized spin-wave theory.
 To understand the roles of constituent atoms on magnetocrystalline anisotropy energy (MAE), we resolve MAE into sublattices and find that a strong negative V-I inter-sublattice contribution is responsible for the relatively small easy-axis MAE in VI$_3$. 
\end{abstract}


\date{\today}
\maketitle

\section{Introduction}
Magnetic 2D van der Waals materials (m2Dv) have great potential for future energy-efficient spin-based devices.
Despite the tremendous advancement in the research field of 2D materials since the discovery of Graphene, finding and developing robust magnetic 2D materials remain a significant challenge.
In practice, it is nontrivial to induce magnetism into non-magnetic 2D materials by the magnetic proximity effects or defect engineering and control them systematically and reliably.
On the other hand, the development of intrinsic magnetic 2D materials have significantly been motivated by the recent experimental breakthrough, demonstrating that magnetism could sustain down to mono-layer $\cri$ and bilayer $\cgt$ at a lower temperature (tens of Kelvins),~\cite{gibertini2019nn, liu2019n, burch2018n, lin2019aami, huang2017n, jiang2018nn, coelho2019jpcc, huang2020nano, linww026, Otrokov2019, soriano2020nanoletter}.
Broader applications require m2Dv with a higher $\tc$, which is essentially determined by their intrinsic magnetic properties, particularly, exchange couplings and magnetocrystalline anisotropy.
The latter is of enhanced importance in 2D as it lifts the constraint of the Mermin-Wagner theorem by inducing a spin-wave (SW) gap and stabilizing the long-range magnetic ordering at finite temperatures.

Besides $\cri$ and $\cgt$, other intrinsic 2D magnetic materials of bulk or layer form have been explored ever since.
Among them, $\vvi$ is a semiconductor with an energy gap of $\sim$\SIrange{0.32}{0.67}{eV}~\cite{son2019prb,kong2019am} in its bulk form.
It is ferromagnetic (FM) below $\SI{50}{K}$ in a relatively small magnetic field of about $\SI{0.1}{T}$~\cite{wilson1987jpcs, tian2019jacs, kong2019am, son2019prb,ming2019}.
$\vvi$ experiences two FM transitions at $\tc \approx$ \SI{36}{K} and \SI{50}{K} at ambient pressure; the two transitions merge into one as hydrostatic pressure above \SI{0.6}{GPa} is applied~\cite{elena2019prb,valenta2020pressure}.
Bulk $\vvi$ has easy-axis magnetic anisotropy ($K_\text{u}$= \SI{37}{kJ/m^{3}} at \SI{10}{K})~\cite{yan2019mae}.
Remarkably, $\vvi$ is a hard ferromagnet with a \SI{9.1}{kOe} coercive field at \SI{2}{K}, significantly larger than in Cr-based m2Dv~\cite{kong2019am, son2019prb}.
In general, coercivity depends not only on magnetic anisotropy but also on extrinsic factors such as microstructures.
The mechanism behind the very different coercivity between $\vvi$ and $\cri$ is not well understood yet.
It is also unclear how different are the exchange couplings and the nature of magnetocrystalline anisotropy (MA) in $\vvi$ compared to those in Cr-based m2Dv.

Unlike CrI$_3$, no exfoliated monolayer $\vvi$ has been reported yet.
However, cleavage energies are theoretically found to be between $0.18$ and $\SI{0.29}{J/m^{2}}$, which is comparable to those for Cr-based m2Dv~\cite{tian2019jacs, ming2019, he2016jmcc, yang2020prb}.
Thus, the exfoliation of $\vvi$ in the near future might be expected, and it is desirable to understand how magnetic interactions evolve with the material configuration changing from bulk to multiple layers, as this understanding is relevant to the real applications, where multiple layers or heterostructures are used instead of bulk materials.

Theoretical studies have also been carried out to understand the intrinsic magnetic properties of monolayer $\vvi$ and conflicting results regarding the exchange couplings and electronic structures have been reported.
For example, large exchange coupling and higher Curie temperature, $\tc$, resulting from a half-metallic ground state, have been reported in Ref.~\cite{yang2020applied}.
On the other hand, Yang and coworkers have found a semiconducting ground state and estimated the energy difference between FM and N\'eel-AFM states, suggesting smaller exchange couplings and $\tc$ in monolayer $\vvi$~\cite{yang2020prb}.
Similarly, different dependences of magnetism on layer stacking in bilayer $\vvi$ have been reported~\cite{wang2020prb, long2020jpclett}.
These discrepancies demonstrate the sensitivity of calculated intrinsic magnetic properties on the underlying electronic structure.

Inelastic neutron scattering (INS) is the tool of choice to characterize the spin-wave spectrum and underlying magnetic interactions.
Unfortunately, unlike $\cri$~\cite{chen2018prx}, no INS study on $\vvi$ has been reported so far.
A comprehensive theoretical analysis of exchange couplings and the resulting spin-wave spectra may provide a helpful insight to understand magnetic properties in $\vvi$ and make predictions upon the prospective INS experiments.

It is believed that MAE in m2Dv results from the interplay between the spin polarization of $3d$ atoms and the large spin-orbit coupling (SOC) of the heavier $p$ elements.
By scaling the SOC strength on Cr and I sublattices, Lado $\etal$ have shown that the heavier I atom plays a significant role in inducing the magnetic anisotropy in $\cri$~\cite{lado20172m}.
On the other hand, Chen $\etal$ have argued that a large single-ion anisotropy is expected in $\vvi$~\cite{yang2020prb} but it is absent in $\cri$.
A numerical resolution of MAE into atomic sites or pairs may help to better understand MAE in $\vvi$.

In this work, we investigate the exchange couplings in monolayer, bilayer, and bulk VI$_3$ within the density functional theory (DFT).
Using the calculated exchange parameters, we employ a spin-wave theory to investigate the spin-wave spectra and estimate the Curie temperatures.
By scaling the SOC strength, we decompose the MAE contribution into sublattices and compare it with $\cri$.
We also discuss the validity of the second-order perturbation theory (SOPT)~\cite{blanco-rey2019njp} in calculating MAE in these systems.

\section{Methods}

The DFT calculations were performed within the framework of the plane-wave projector-augmented wave formalism~\cite{blochl1994prb}, as implemented in the Vienna $\abinitio$ simulation package (\textsc{vasp})~\cite{kresse1999prb,kresse1996cms}.
The Perdew-Burke-Ernzerhof generalized gradient approximation~\cite{perdew1996prl} was used for the exchange-correlation functional.
All calculations were performed with a plane wave cutoff energy of $\SI{400}{eV}$.
To properly describe the additional Coulomb repulsion between electrons beyond DFT, the on-site Hubbard $U$ was applied on V-$3d$ electrons using the double-counting scheme introduced by Dudarev $\etal$~\cite{dudarev1998prb}.

The experimental~\cite{tian2019jacs} lattice parameters and atomic positions of bulk $\vvi$ were employed for all calculations unless explicitly stated otherwise.
For bulk $\vvi$, we confirmed that the volume of the optimized structure is within $\sim4\%$ of the experimental value.
The reported optimized lattice constants for monolayer $\vvi$ are different only by $\sim3\%$ from the experimental values of bulk $\vvi$~\cite{ming2019,Subhan2020}.
Thus, employing experimental structural parameters is a reasonable approximation for this study.
A more comprehensive investigation of theoretical structural optimization, which often depends on the functionals used, is beyond the scope of the current study.
In the bilayer and monolayer models, a vacuum region of 16\AA was included to avoid the interaction between periodic images.

\subsection{Exchange parameters}

\begin{figure}[htb]
  \centering
  \includegraphics[width = 0.8\linewidth]{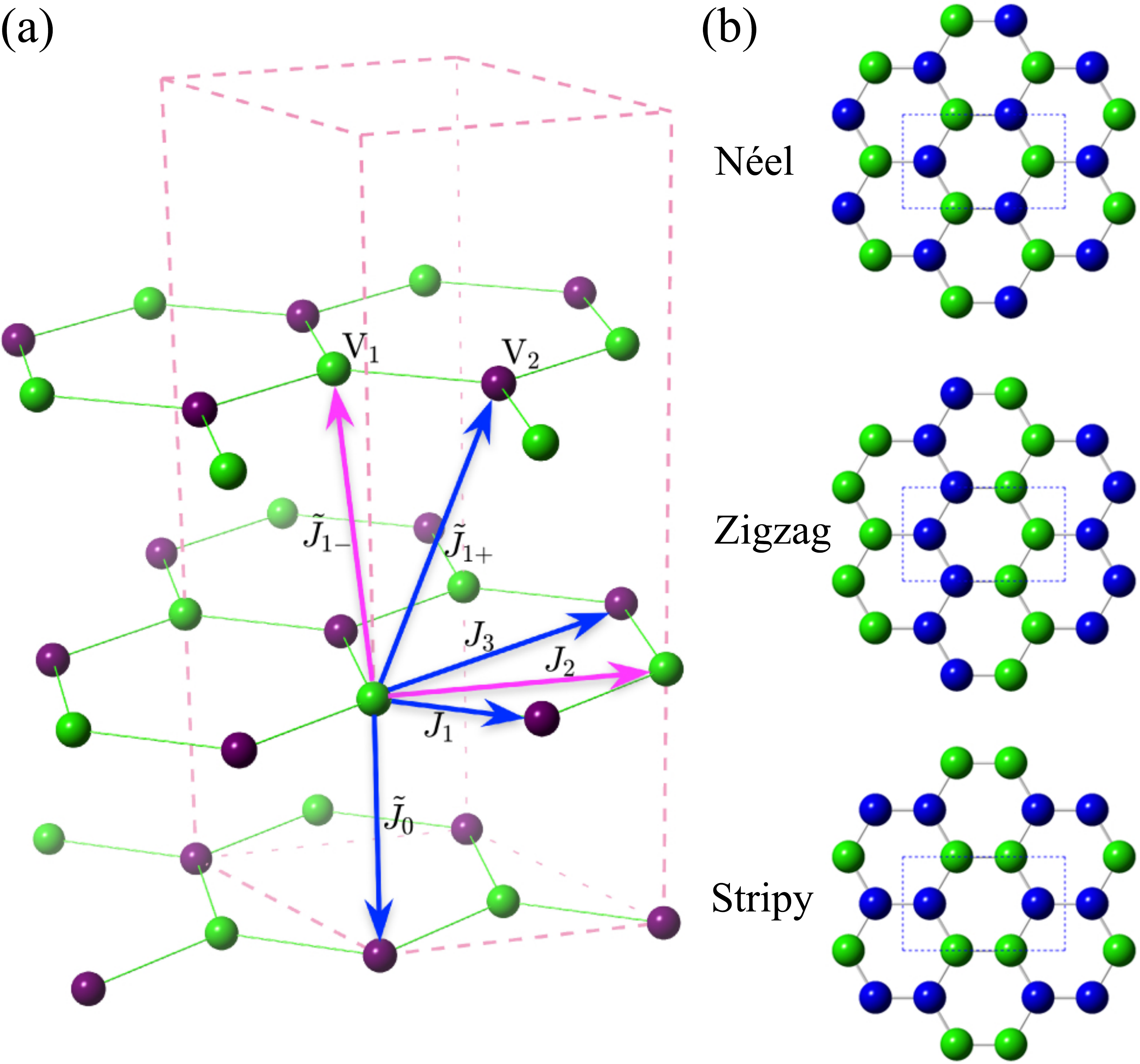}
  \caption{ (a) Schematic representation of the V sublattices in rhombohedral $\vvi$. The hexagonal conventional unit cell is used to depict pair exchange parameters for the first few neighbors of V atoms. Two V sublattices, V$_1$ and V$_2$, are indicated with green and purple spheres, respectively. (b) N\'eel-, zigzag-, and stripy-AFM intralayer spin configurations. A four-atom unit cell, denoted by the dashed blue lines, is used to describe the three spin configurations. Green and blue spheres denote the spin-up and spin-down orientations, respectively.}
  \label{fig:xtal}
\end{figure}

To estimate the exchange couplings, we map the total energies of various collinear spin configurations onto a Heisenberg model defined as

\begin{equation}
  H=-\sum_{i\neq j} J_{ij}\, \hat{\bf e}_i\cdot\hat{\bf e}_j,
  \label{eq:Hamiltonian_e}
\end{equation}
where $\hat{\bf e}_i$ is the unit vector of magnetic moment on site $i$.
The first six nearest-neighbor exchange couplings $J_{ij}$, including three intralyer ($J_1$, $J_2$, and $J_3$) and three interlayer ($\tj_0$, $\tj_{1-}$, and $\tj_{1+}$) exchange couplings indicated in \rfig{fig:xtal}(a), are included to describe the bilayer and bulk cases.

For monolayer $\vvi$, we calculate the energies of four collinear spin configurations, namely, FM, N\'eel-, zigzag-, and stripy-AF~\cite{sivadas2015spinconfig} to derive the three intralayer exchanges as follows:
\begin{equation}
  \left(
  \begin{array}{c}
    J_1 \\
    J_2 \\
    J_3 \\
  \end{array}
  \right)
  =
  \left(
  \begin{array}{ccc}
    6 &     0 &     6  \\
    2 &     8 &     6  \\
    4 &     8 &     0  \\
  \end{array}
  \right)^{-1}
  \left(
  \begin{array}{c}
    \Delta E_\text{N} \\
    \Delta E_\text{Z} \\
    \Delta E_\text{S} \\
  \end{array}
  \right).
  \label{eq:ec}
\end{equation}
Here, $\Delta E_\mu = E_\mu -E_\text{FM}$, $\mu=$N, Z, and S, are the energies (per V atom) of N\'eel, Zigzag, and Stripy configurations (illustrated in \rfig{fig:xtal}(b)) with respect to the reference FM configuration.

To obtain the interlayer couplings for bilayer and bulk $\vvi$, we double each of these four intralayer configurations along the out-of-plane direction in a larger supercell.
For each intralayer configuration, there are two (analog to FM/AFM) possible interlayer orderings, whose energy difference is used to extract the interlayer couplings.

\subsection{Linear spin-wave theory}
Using the obtained exchange parameters we calculate the SW spectra within a linear SW theory.
First, we rewrite \req{eq:Hamiltonian_e} in terms of spin ${\bf S}$ instead of unit vector $\hat{\bf e}$,  as $H=- \frac12 \sum_{i\neq j} J_{ij}^{S}\, {\bf S}_i\cdot{\bf S}_j$, where
$J_{ij}^{S}=2J_{ij}/S^2$, and $S=|{\bf S}|=m/2$ is the on-site spin of V atoms.
Next, we bosonize the spin operator via the Holstein-Primakoff transformation~\cite{HolPrim1940} and truncate the Hamiltonian at quadratic order in bosons to diagonalize it.
There are two V sublattices in the primitive unit cell of bulk R-$\vvi$, resulting in two SW branches with energies written as~\cite{ke2021ncm}:

\begin{equation}
  \label{eq:sw}
  \omega_\pm(\bfq)= S \left( J_{\bfq=0}^\text{Intra} + J_{\bfq=0}^\text{Inter}
  - J_{\bfq}^\text{Intra} \pm | J_{\bfq}^\text{Inter} | \right),
\end{equation}
where $J_{\bfq}^\text{Intra}$ and $J_{\bfq}^\text{Inter}$ sum over the intra-sublattice and inter-sublattice exchange couplings, respectively.
With the considered range of exchange couplings, we have
\begin{eqnarray}
\label{ABq}
J_{\bfq}^\text{Intra} &=& J_2(\bfq)+\tj_{1-}(\bfq), \nonumber \\
J_{\bfq}^\text{Inter} &=& J_1(\bfq) + J_3(\bfq)  + \tj_{0}(\bfq) + \tj_{1+}(\bfq).
\end{eqnarray}
Here,
\begin{equation}
\label{Jq}
 J_i(\bfq)= J_i^S  \gamma_i(\bfq) = J_i^S  \sum_{\boldsymbol{\delta}_i} e^{-i 2 \pi \bfq\cdot \boldsymbol{\delta}_i },
\end{equation}
and $\gamma_i(\bfq)$ is the structure factor, while $\boldsymbol{\delta}_i$ standing for the  connecting vectors of corresponding exchange $J_i^S$.

\subsection{Non-linear spin-wave theory}
The linear SW theory briefly discussed above neglects the interactions between magnons.
Therefore it is applicable only at the lower temperatures, where INS experiments are often used to measure SW spectra.
In the critical region close to the Curie temperature, interactions between the spin waves become progressively more relevant.
To the leading order, these interactions are given by the quartic terms of the Holstein-Primakoff transformed Hamiltonian.
To assess the SW interaction effects, we involve quartic terms and treat them within a Hartree-Fock-like decoupling approximation~\cite{bloch1962prl, sarker1989prb, liu1992jpcm, li2018jmm}.
This procedure leads to an effective quadratic Hamiltonian containing renormalization factors that encapsulate SW interaction effects.
The effective quadratic Hamiltonian is then solved self-consistently, e.g., for the magnetization versus temperature behavior, $M(T)$.
In such a way, we set up a self-consistently renormalized spin-wave theory (SRSWT)~\cite{mkhitaryan2021srswt}, which is the analogue of previous self-consistently renormalized theories~\cite{bloch1962prl, loly1971jpcs, rastelli1974jpcs, pini1981jpcs} for layered systems with hexagonal intralayer structure.

Within our SRSWT, renormalization factors defining the effective Hamiltonian are combinations of the average spin polarization, $\bar{S}$, and the average short-range two-boson correlations between the exchange-coupled sites.
Thus, the effective Hamiltonian is a function of four parameters in monolayer $\vvi$ or of seven parameters in bulk $\vvi$.
The self-consistency equations are found by deriving relations for $\bar{S}$ and the average two-boson correlations from the effective Hamiltonian.
Subsequently, these equations are solved numerically by successive iterations.
The magnetization is defined as the normalized average spin polarization, $M(T)=\bar{S}(T)/S$, and the critical temperature is taken as the temperature at which $M(T)$ turns to zero.

Using the calculated values of exchange couplings, we find $M(T)$ and the corresponding critical temperatures for monolayer, bilayer, and bulk VI$_3$, by means of SRSWT.
The analysis of monolayer and bulk systems is facilitated by the fact that these systems consist of two equivalent sublattices of magnetic atoms.
In contrast, we regard the few-layer systems with $L$ number of layers as consisting of $2L$ sublattices -- two per hexagonal layer.
This approach allows us to take into account the physical difference between the surface and bulk layers.
At the same time, it requires a substantially larger parameter space for the self-consistency equations, as the average spin polarization and two-boson correlation values at different sublattices are generally different.
The details of our SRSWT method and implementation can be found elsewhere~\cite{mkhitaryan2021srswt}.

\subsection{Sublattice-resolved MAE}
MAE is calculated as the total-energy difference between two FM states, in which the magnetization is aligned along the [100] or [001] directions, respectively.
SOC was included using the second-variation method~\cite{li1990prb,shick1997prb}.
We resolve MAE contributions into sublattices via scaling the SOC strength $\xi_i$ on V and I sites with parameters $\lambda_i$;
the SOC Hamiltonian becomes
\begin{equation}\label{eq:hamilsoc}
H_{\text{soc}} = \sum_{i=\text{V,I}} \lambda_i \xi_i ({\bf L}_{i} \cdot {\bf S}_{i}).
\end{equation}
Assuming that MAE can be well described by SOPT, total MAE can be written as ~\cite{ke2019prb}
\begin{equation}\label{eq:secorpergen}
  K(\lambda_\text{V},\lambda_\text{I}) = K_\text{V-V}{\lambda_{\text{V}}^2} + K_\text{V-I}{\lambda_{\text{V}}}{\lambda_{\text{I}}} + K_\text{I-I}{\lambda_{\text{I}}^2},
\end{equation}
where $K_\text{V-V}$, $K_\text{V-I}$, and $K_\text{I-I}$ describe the response of MAE to the SOC strength of V, V and I simultaneously, and I, respectively.
They can be determined by fitting the $K(\lambda_\text{V},\lambda_\text{I})$ curve or directly evaluated within SOPT, and the inter-sublattice term $K_\text{V-I}$ arise from the coupling between the two sublattices~\cite{ke2019prb}.

\section{RESULTS AND DISCUSSION}

\subsection{Electronic properties}
\begin{figure*}[bht]
    \centering
    \includegraphics[width =1.0 \linewidth]{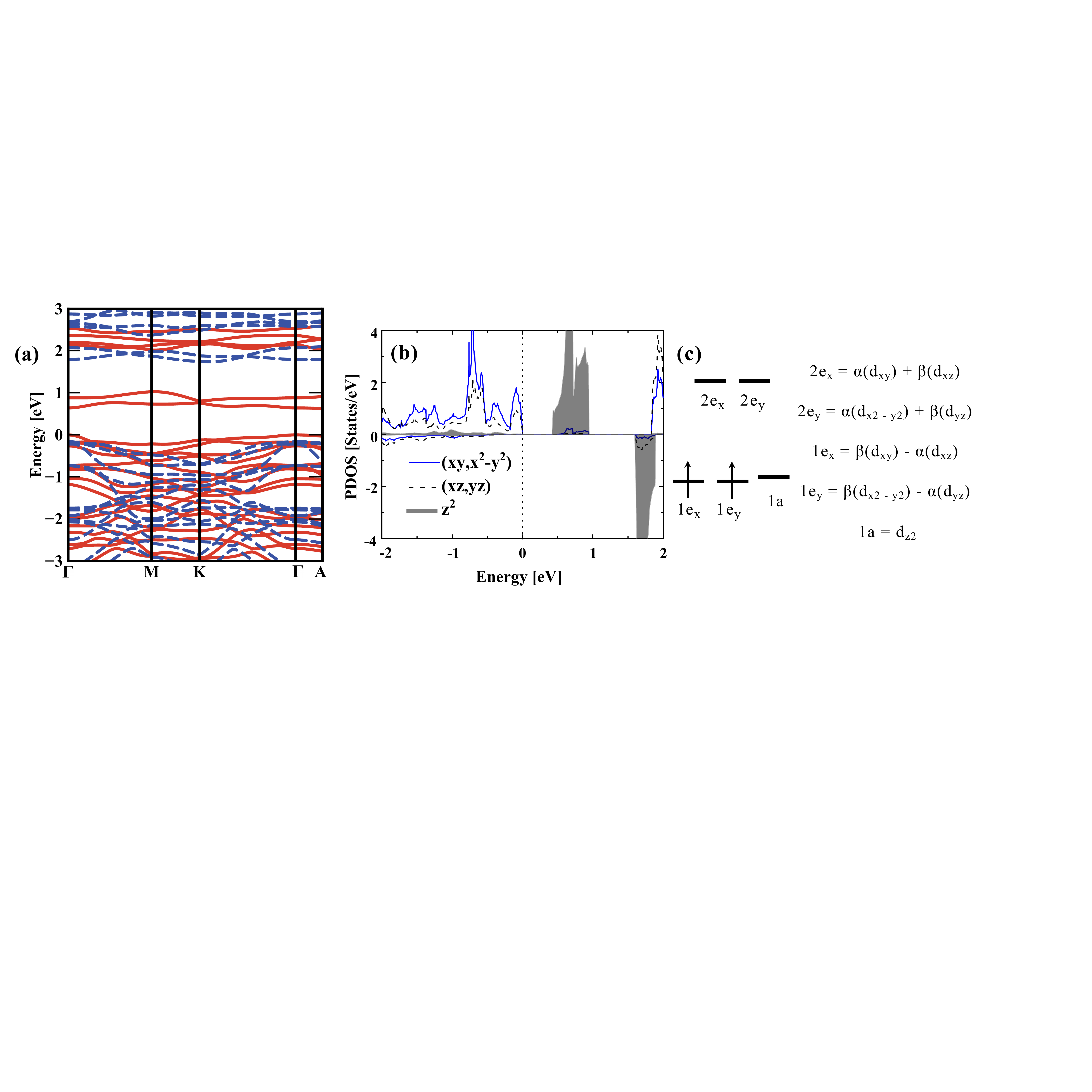}
    \caption{(a) Band structure of $\vvi$ for $U=$ \SI{2}{eV}.
        Solid and dotted lines indicate the majority and minority spin channels, respectively.
        Fermi level $\ef$ is set at zero energy. This figure was constructed using \textsc{sumo}~\cite{ganose2018sumo}.
        (b) Partial density of states projected on the V-$d$ states. (c) The splitting and occupation of V-$d$ orbitals in $\vvi$, where $\alpha=\sqrt{1/3}$ and $\beta=\sqrt{2/3}$; the $t_{2g}$ orbital set splits into doubly-degenerate ($1e_x$, $1e_y$) and singly-degenerate ($1a$) orbital sets.}
    \label{bands}
\end{figure*}

\rFig{bands} shows the calculated non-SOC band structures, the partial density of states (PDOS) projected on V-$d$ orbitals, and their corresponding orbital characters near the Fermi level $\ef$, calculated in DFT+$U$ with $\ueff= \SI{2}{eV}$.
The obtained bandgap of $\SI{0.42}{eV}$ is within the range of reported experimental values~\cite{kong2019am, son2019prb}.

The oxidation state of V in $\vvi$ is $+3$, with electronic configuration [Ar]4$s^03d^2$.
V atoms arrange in a honeycomb structure within the layer with edge-sharing octahedral coordination by I ligands.
The octahedral ligand and crystal fields split the V-$d$ states into the doublet $e_g$ and triplet $t_{2g}$ subsets, with the latter states lying lower in energy.
Therefore, the two unpaired electrons of V$^{3+}$ occupy the $t_{2g}$ orbitals, both with spins up according to Hund's first rule, forming the spin state $S=1$, consistent with the calculated on-site magnetic moment of V atom $m\approx 2 \mub$/V.

Correlation effects beyond DFT are required to describe the semiconducting ground state of $\vvi$ correctly~\cite{lee2020prbr}.
\rFig{bands}(c) shows that the $t_{2g}$ orbital set further splits into doubly degenerate ($1e_x$, $1e_y$) and singly degenerate ($1a$) orbital sets.
While DFT gives a half-metallic state, the on-site Coulomb correction shifts the $1a$ state upward relative to the ($1e_x$, $1e_y$) states, resulting in a bandgap in between~\cite{lee2020prbr, wang2020prb}.
We note that one may need to control the initial orbital occupancy in the DFT+$U$ calculation to converge to the experimental semiconducting ground state for  $\vvi$, which may explain the discrepancy of ground states found in previous DFT+$U$ calculations~\cite{wang2020prb,yang2020prb,huang2020pccp}.

We also qualitatively explore the effects of compressive strain on the electronic structure by gradually decreasing the lattice parameters (volume) of bulk $\vvi$ while keeping the aspect ratio fixed.
We found that the ($1e_x$, $1e_y$) states shift upward relative to the $1a$ state, and a semiconductor-to-metal transition occurs at a strain of 5–6\% when the gap closes, and all three $t_{2g}$ states become partially occupied.
A more systematic and comprehensive theoretical investigation on structural properties may help compare with or guide the future experiments in this direction.

\subsection{Exchange coupling in monolayer, bilayer, and bulk $\vvi$}
\label{section:magnetostructuralcoupling}

\begin{table}[bht]
  \caption{Pairwise intralayer exchange parameters $J_i$, $i=1$, $2$,
    $3$ and interlayer exchange parameters $\tj_{i}$, $i=0$, $1+$, $1-$
for the Heisenberg Hamiltonian $H=-\sum_{i\neq j} J_{ij}\, \hat{\bf e}_i\cdot\hat{\bf e}_j$
in monolayer, bilayer, and bulk R-$\vvi$ calculated within DFT+$U$ with various $U$ values (in the unit of eV).
Positive (negative) $J_{ij}$ values correspond to FM (AFM) couplings.
Exchange parameters are illustrated in Fig.~\ref{fig:xtal} for the bulk structure.
The degeneracy (No.) and distance ($R_{ij}$) of $J_{ij}$ are also provided.
  }
  \bgroup \def\arraystretch{1.05}%
\begin{tabular*}{\linewidth}{l@{\extracolsep{\fill}}lllcccrrr}
  \hline  \hline
  \\[-1em]
  R-$\vvi$  & &  \multirow{2}{*}{Lbl.} & \multirow{2}{*}{No.} && $R_{ij}$  && \multicolumn{3}{c}{$J_{ij}$(meV)}\\
  \\[-1.1em]
  \cline{1-1} \cline{6-6}  \cline{8-10}
  \\[-1.1em]
Form &  &   &  & & (\AA) & & $U$=2 & $U$=3 & $U$=3.7 \\
  \\[-1.1em]
  \hline
  \\[-1.1em] \multirow{3}{4.5em}{Monolayer}
& & $J_1$         & 3                   & & 3.946 &&  2.26  &   2.07 &   1.99 \\
& & $J_2$\footnotemark[1]         & 6\footnotemark[1]   & & 6.835 &&  0.06  &   0.08 &   0.09 \\
& & $J_3$         & 3                   & & 7.893 && -0.23  &  -0.18 &  -0.16 \\    \hline \multirow{7}{3em}{Bilayer}
& & $J_1$         & 3                   & & 3.946 &&  2.26  &   2.07 &   2.03 \\
& & $J_2$         & 6                   & & 6.835 &&  0.04  &   0.06 &   0.06 \\
& & $J_3$         & 3                   & & 7.893 && -0.16  &  -0.14 &  -0.12 \\
& & ${\tj_{0}}$   & 1/0\footnotemark[2] & & 6.552 && -0.14  &  -0.07 &  -0.08 \\
& & ${\tj_{1-}}$  & 3                   & & 7.660 &&  0.08  &   0.06 &   0.04 \\
& & ${\tj_{1+}}$  & 0/3\footnotemark[2] & & 7.672 &&  0.25  &   0.17 &   0.18 \\    \hline \multirow{7}{2em}{Bulk}
& & $J_1$         & 3                   & & 3.946 &&  2.18  &   2.02 &   1.98 \\
& & $J_2$\footnotemark[1]         & 6   & & 6.835 &&  0.10  &   0.09 &   0.08 \\
& & $J_3$         & 3 & & 7.893 && -0.20    &  -0.16 &  -0.15 \\
& & ${\tj_{0}}$   & 1 & & 6.552 &&  0.27    &   0.17 &   0.14 \\
& & ${\tj_{1-}}$\footnotemark[1] & 6 & & 7.660 &&  0.18 &   0.13 &   0.11 \\
& & ${\tj_{1+}}$  & 3 & & 7.672 && -0.05    &  -0.03 &  -0.02 \\
  \hline\hline
\end{tabular*}
\egroup
\footnotetext[1]{Intra-sublattice couplings for monolayer and bulk cases.}
\footnotetext[2]{Different V sites have different numbers of neighbors coupled through $\tilde{J}_0$ or $\tilde{J}_{1+}$.}
\label{Tab:exchangeint}
\end{table}

\begin{figure}[htb]
  \centering
  \includegraphics[width =0.98 \linewidth]{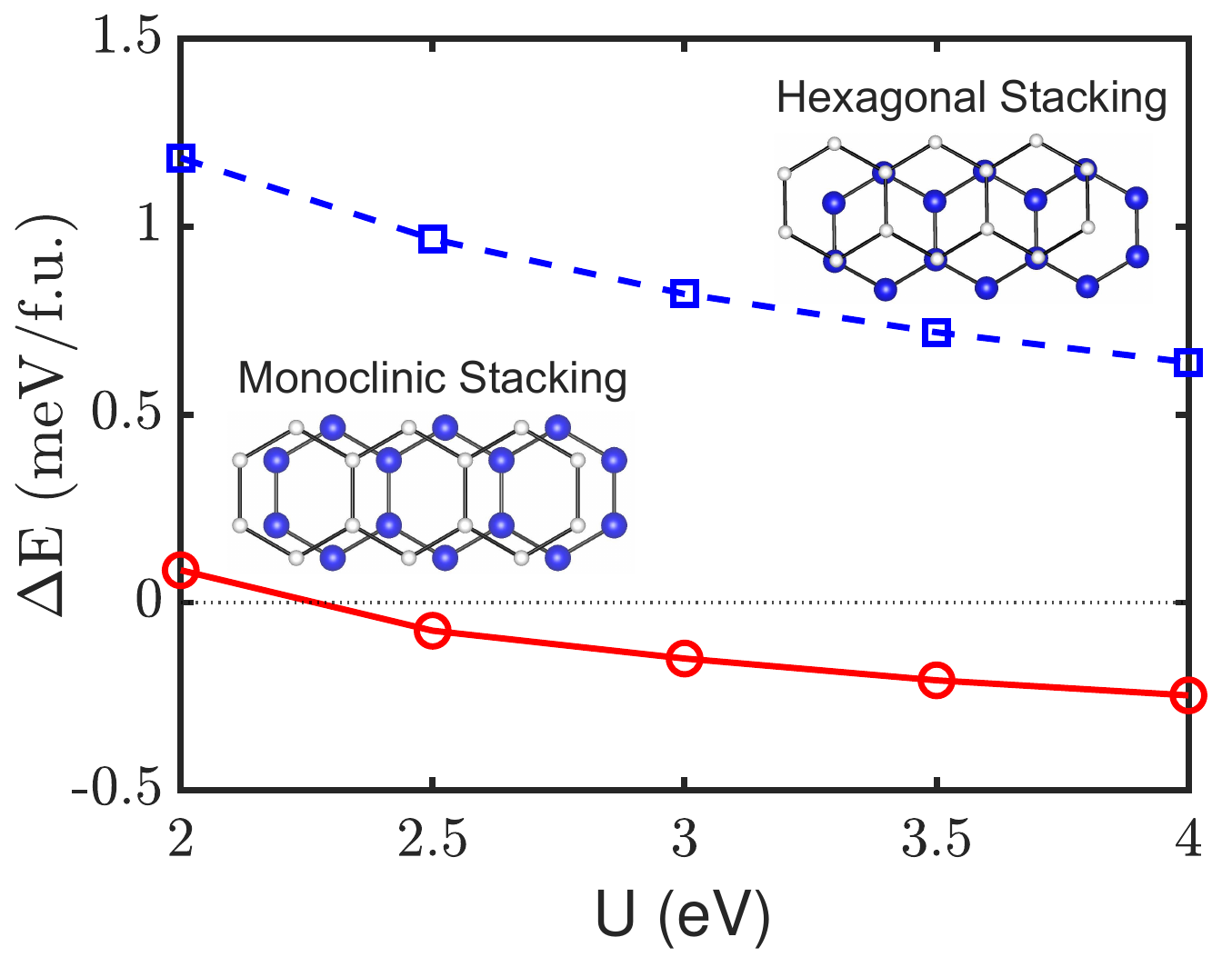}
  \caption{ Interlayer magnetic oupling energy, $\Delta E= E_\text{AFM}^\text{L-L} - E_\text{FM}^\text{L-L}$, as a function of on-site Coulomb interaction parameter $\ueff$ in bilayer $\vvi$ with hexagonal and monoclinic stackings.
}
  \label{bilayerenergy}
\end{figure}

We next discuss the dependence of exchange couplings on dimensionality, additional on-site Coulomb interaction $U$, and stacking order.
\rtbl{Tab:exchangeint} summarizes the exchange parameters calculated in monolayer, bilayer, and bulk $\vvi$ with different $U$ values.
Although increasing $\ueff$ values generally decrease the exchange coupling, suggesting a more localized-moment picture, qualitative details remain the same.
In particular, there is no change in their signs for the considered $\ueff$ values.

In all the monolayer, bilayer, and bulk forms, the nearest neighbor intralayer exchange coupling $J_1$, arises from a superexchange via the near-90$^\circ$ V-I-V bonds. It is found to be dominant and FM, while $J_2$ and $J_3$ are weakly FM and AFM, respectively.
In bulk $\vvi$, the first two nearest neighbor interlayer parameters $\tj_{0}$ and $\tj_{1-}$ are found to be positive, leading to a favorable FM interlayer ordering.
Note that although $\tj_{1+}$ is negative, its magnitude is small.
The overall interlayer coupling is dominated by $\tj_{0}$ and $\tj_{1-}$, especially the latter, which has a larger coordination number of six.

Interestingly, the interlayer couplings show a much stronger dependence on dimensionality than the intralayer ones.
As shown in \rtbl{Tab:exchangeint}, the intralayer couplings only change slightly in three different forms.
In contrast, the overall interlayer coupling in bilayer remains FM as in bulk, the individual interlayer couplings $\tj_0$ and $\tj_{1+}$ change signs when we compare the bilayer and bulk cases.
Note that the coordination number of interlayer couplings decreases by a factor of two in the bilayer form.

We also investigate the stacking-dependent magnetism in $\vvi$ and found a similar magnetostructural coupling as previously found in $\cri$~\cite{sivadas2018nl,sariano2019ssc,jiang2019prb}.
\rFig{bilayerenergy} shows the $\ueff$ dependence of interlayer magnetic coupling energy in bilayer $\vvi$ with hexagonal stacking and monoclinic stacking.
The energy difference is calculated as $\Delta E= E_\text{AFM}^\text{L-L} - E_\text{FM}^\text{L-L}$, with $E_\text{AFM}^\text{L-L}$ and $E_\text{FM}^\text{L-L}$ being the energies of AFM-ordered layer-layer and FM-ordered layer-layer configurations, respectively.
For the hexagonal stacking, FM interlayer ordering between the layers is found to be stable over a wide range of $\ueff$.
For the monoclinic stacking, in contrast, the energy difference $\Delta E$ is very small at $\ueff\approx\SI{2}{eV}$, indicating competing interlayer AFM and FM states.
Moreover, a crossover from the interlayer-FM to interlayer-AFM magnetic ground state occurs at $\ueff\approx\SI{2.25}{eV}$.
These results point to the strong magnetostructural coupling in bilayer $\vvi$.
Consequently, control of the magnetic configuration, for example, via strain or electric field, is feasible, facilitating device applications.

\subsection{Spin-wave dispersion}
\begin{figure}[htb]
  \centering
\begin{tabular}{c}
\includegraphics[width = 0.95\linewidth]{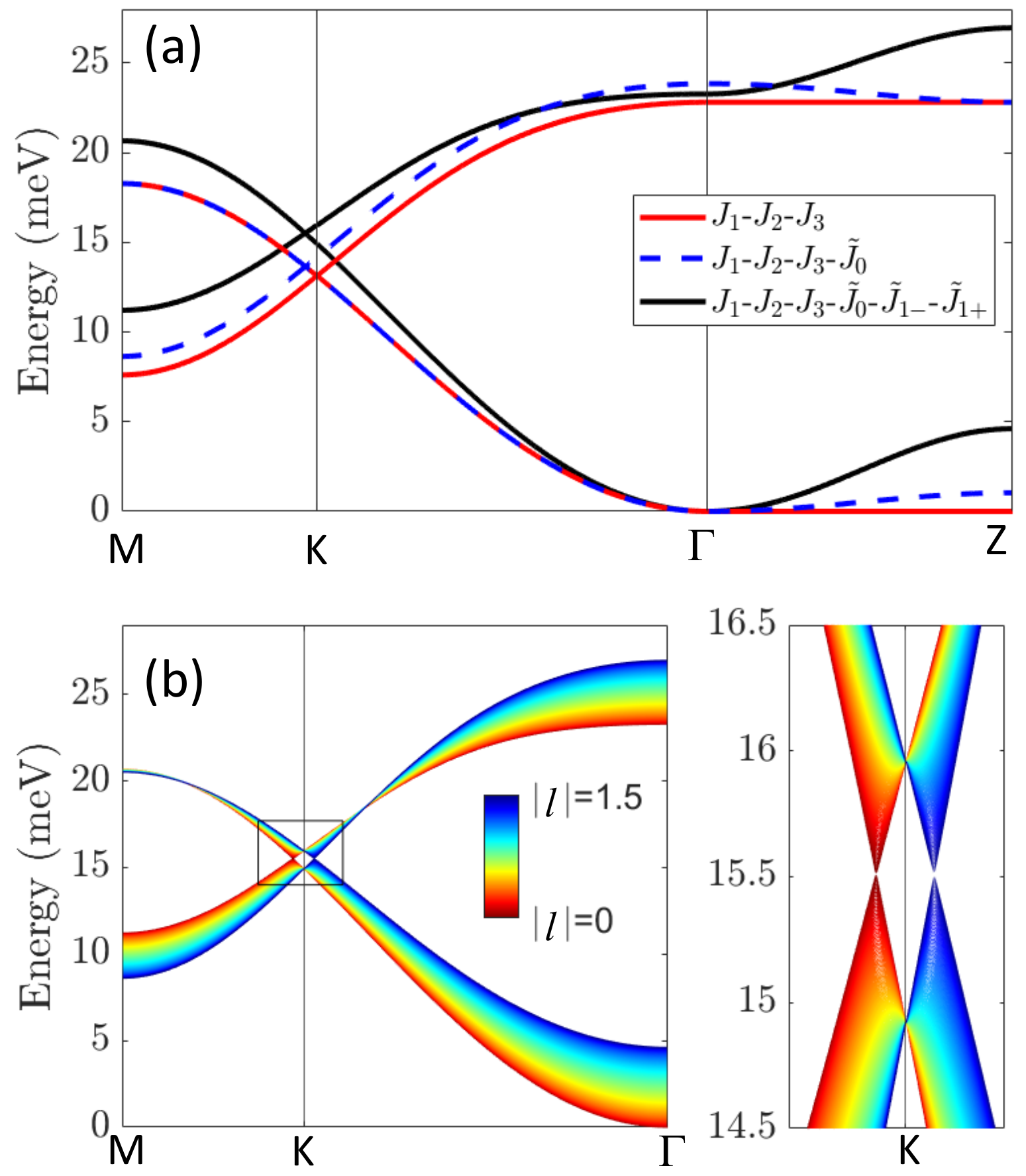}
\end{tabular}%
  \caption{ (a) Spin-wave spectra calculated with only intralayer exchange couplings $J_i$, $i=1$, $2$, $3$ involved (red), intralayer $J_i$, $i=1$, $2$, $3$, and interlayer $\tj_0$ involved (blue dashed), and all calculated $J_i$ and $\tj_i$ exchange couplings involved (black). For all three plots, the exchange coupling values used are those calculated for the bulk $\vvi$.  (b) Spin-wave spectra calculated along [$h$,$k$,$l$] with various $l$ ($k_z$) values; [$h$,$k$,0] is along the $\Gamma$--$K$--$M$ path (i.e., it is equivalent to the red line in panel a). }
\label{fig:spinwave}
\end{figure}

Utilizing the exchange parameters for the bulk $\vvi$, we calculated the linear SW spectra and investigated how different exchanges affect the spectra features.
Such understanding may be useful to compare with future INS experiments on $\vvi$.
\rFig{fig:spinwave}(a) shows the SW spectra calculated along the in-plane $\Gamma$--$K$--$M$ path and out-of-plane $\Gamma$--$Z$ path.
To better illustrate the effects of different exchanges on SW excitation, we consider three cases with different numbers of interlayer couplings included in the Hamiltonian: 1) only intralayer couplings ($J_1$-$J_2$-$J_3$; solid red lines), 2) intralayer and the $1st$ nearest-neighbor interlayer couplings ($J_1$-$J_2$-$J_3$-$\tj_{0}$; blue dashed lines), 3) all of the considered intralayer and interlayer couplings ($J_1$-$J_2$-$J_3$-$\tj_{0}$-$\tj_{1-}$-$\tj_{1+}$; solid black lines).

With only intralayer coupling considered, two SW branches cross exactly at $K$, which is caused by the vanishing of the inter-sublattice coupling $J_{\bfq}^\text{Inter}$, and thus the energy difference between in-phase and out-of-phase inter-sublattice precessing.
Including the first nearest interlayer coupling $\tj_0$ leaves the acoustic branch unchanged but lifts the optical branch by a constant and correspondingly shifts the Dirac crossing toward $M$ point along the $\Gamma$--$K$--$M$ path.
The constant shifting of in-plane optical SW spectra is because $\tj_0$ is along the $z$ direction.
On the other hand, the SW spectra along the $\Gamma$--$Z$ direction become dispersive with $\tj_0$.
Unlike $\tj_0$, interlayer couplings $\tilde{J}_{1+}$ and $\tilde{J}_{1-}$ have nonzero in-plane components in their connecting vectors (similar to those of $J_1$).
They induce additional dispersion on in-plane SW spectra but preserve the Dirac crossing at $K$ (not shown) in the absence of $\tj_0$.
We note that comparing the dispersions of the acoustic and optical modes along the $\Gamma$--$Z$ direction may tell whether the dominant FM interlayer coupling is intra-sublattice or inter-sublattice.
For example, as shown in \rfig{fig:spinwave}(a), FM $\tj_{1-}$ increases both branches' energies while the FM $\tj_0$ increases the acoustic mode energy but lowers the optical mode energy when we move along $\Gamma$--$Z$.
It would be interesting if future INS can be used to unveil the nature of interlayer couplings.

\rFig{fig:spinwave}(b) shows the SW spectra in bulk $\vvi$ calculated along paths that are parallel to $\Gamma$--$K$--$M$  but at various $k_z$ planes (denoted by different colors in \rfig{fig:spinwave}(b)). All of the six exchange couplings are included.
At finite $k_z$, the interlayer couplings rotate the Dirac crossing off this path.
Even at $k_z=0$ plane, including further interlayer couplings can also rotate the Dirac crossing off the high-symmetry line and opens up a gap along the $\Gamma$--$K$--$M$ high-symmetry path~\cite{ke2021ncm}.

Relativistic exchanges and MAE are not included in the model spin Hamiltonian to calculate SW spectra.
Easy-axis MAE, in either single-ion or anisotropic exchange (equivalently two-ion MAE) form, introduces a gap of $\Delta=(2S-1)K$ at $\Gamma$.
On the other hand, Dzyaloshinkii-Moriya interactions (DMI) and Kitaev interaction can open up a global gap~\cite{chen2018prx} between two magnon branches.

\subsection{Magnetization vs Temperature from non-linear spin-wave theory}

\begin{figure}[t] \centerline{\includegraphics[width = 0.95\linewidth,angle=0,clip]{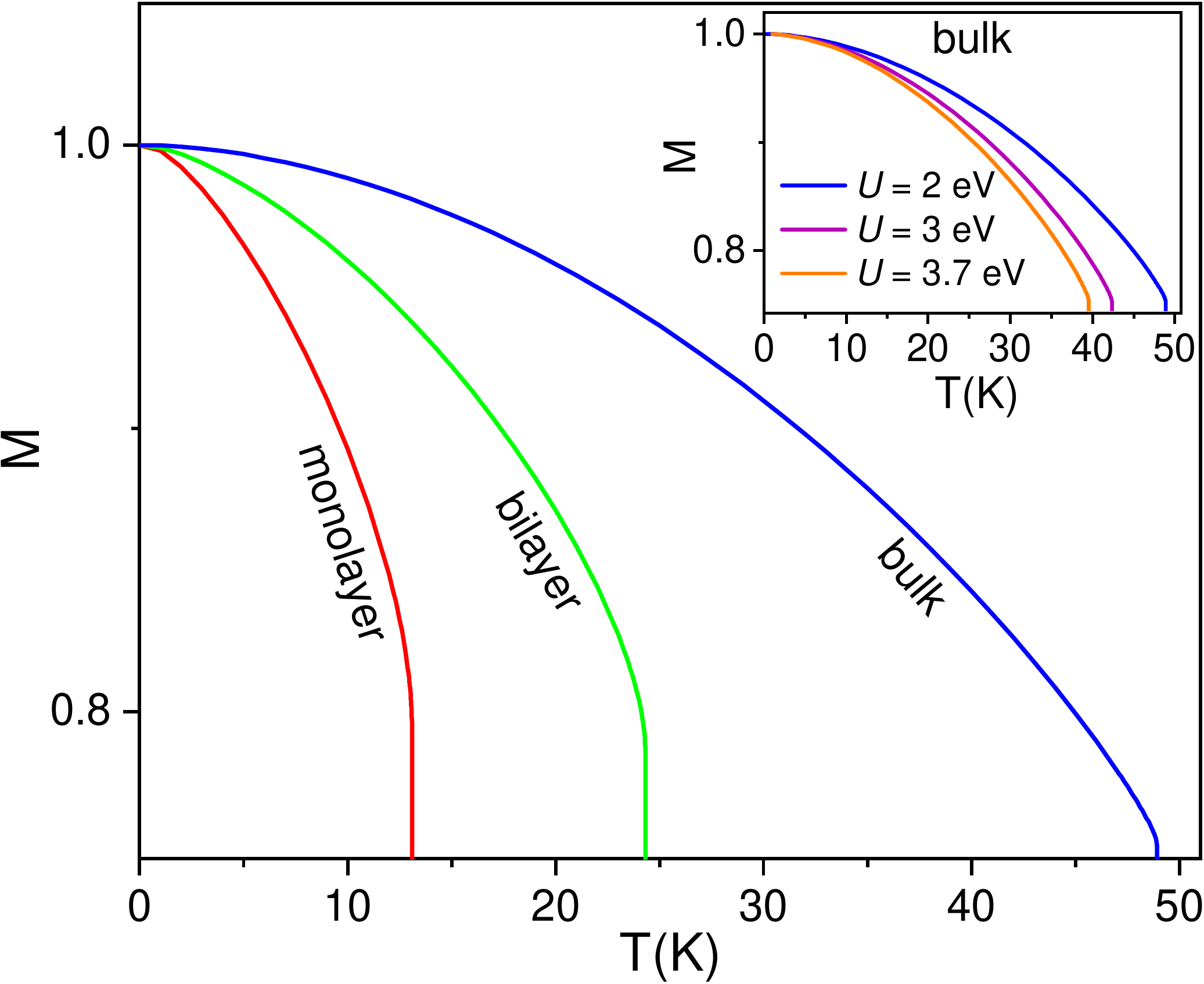}}
\caption{(Color online) Magnetization dependence on temperature in
  zero magnetic field calculated using self-consistently renormalized spin-wave theory~\cite{mkhitaryan2021srswt}.
Plots for the monolayer, bilayer, and bulk VI$_3$ are obtained by using exchange parameters calculated for $U=2$~eV, as listed in \rtbl{Tab:exchangeint}, and the calculated MAE value of 71~$\mu$eV.
Inset: the same dependence for the bulk system, resulting from the three different sets of exchange couplings calculated for $U=2$, $3$, and $3.7$~eV.}
\label{srswtfig}
\end{figure}

We further use SRSWT to explore the temperature dependence of magnetization, $M(T)$, of the monolayer, bilayer, and bulk VI$_3$, by utilizing the calculated MAE and exchange parameters.
For the bulk system, exchange parameters calculated for on-site Coulomb correction of 2, 3, and 3.7~eV yield magnetic ordering temperature values of 48.9, 42.3, and 39.5~K, respectively (see Fig.~\ref{srswtfig} inset).
By noting that the ordering temperature found for $U=2$~eV is in excellent agreement with the experimentally reported value of 49~K~\cite{kong2019am}, for the analysis of monolayer and bilayer systems, we use the exchange couplings calculated with this value of $U$.
For the momentum-space integrals involved in the SRSWT self-consistency equations, we use a very fine $k$-point mesh of 100$\times$100 for 2D and 100$\times$100$\times$100 for 3D.

The plots in \rfig{srswtfig} show a typical mean-field-like behavior, implying a first-order phase transition with the magnetization abruptly vanishing at a critical temperature.
As expected, the magnetization is strongly quenched in lower dimensions;
The ordering temperature decreases to about 13~K and 24.3~K in monolayer and bilayer $\vvi$, respectively.
Overall, these results attest to the feasibility of intrinsically ferromagnetic monolayer and bilayer VI$_3$.

\subsection{Constituents of MAE and validity of perturbation treatment}
\begin{figure}[htb]
    \centering
    \includegraphics[width = .95\linewidth]{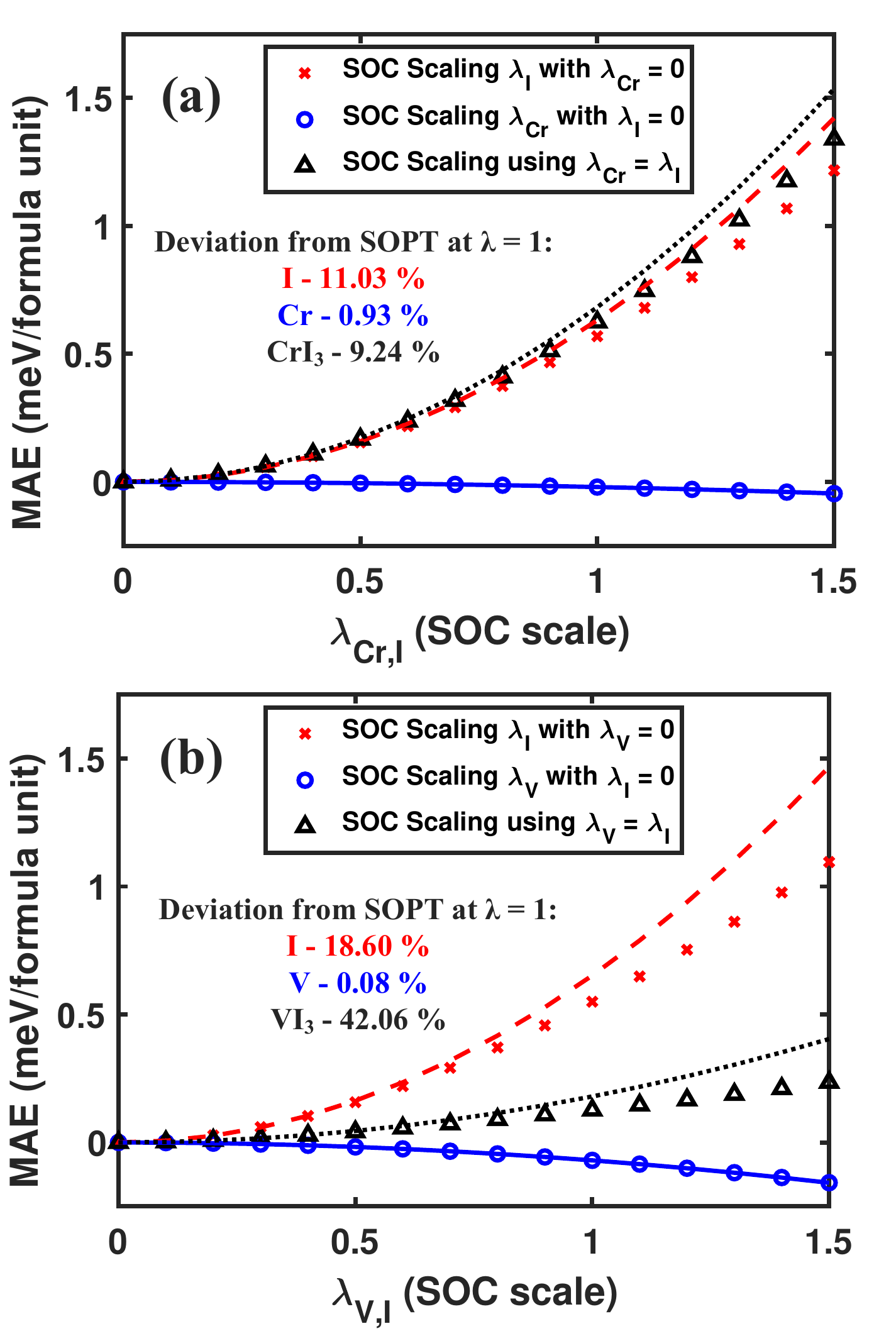}
    \caption{MAE as a function of the SOC scaling factor $\lambda_{M,\text{I}}$ ($M$ = V or Cr) in monolayer $\cri$ (a) and $\vvi$ (b).
The calculated values are presented by symbols, and the lines are fits using the data calculated with $\lambda\in[0,  0.4]$. 
The red dashed, blue straight, and black dotted lines correspond to scaling with $\lambda_{\text{I}}$ (at $\lambda_{M} = 0$), $\lambda_{M}$ (at $\lambda_{\text{I}} = 0$), and $\lambda_{\text{I}}=\lambda_{M}$, respectively. The calculated MAE at $\lambda_{\text{I}}=\lambda_{M}=1$ corresponds to normal DFT result.}
    \label{scale}
\end{figure}

The calculated MAE in $\vvi$ is much smaller than in $\cri$.
To understand the role of constituent atoms on the MA in two compounds we vary the SOC strength on V/Cr and I sites independently or simultaneously, to study how MA evolves in monolayer $\vvi$ and compared it with monolayer $\cri$.
\rFig{scale} shows the calculated MAE as functions of scaling parameter $\lambda$ with three different scaling schemes, as denoted by red cross, blue triangle, and blue circles.
We extract corresponding sublattice contributions by fitting \req{eq:secorpergen} data calculated with $\lambda\in[0,  0.4]$.

The large SOC on I sites is essential in providing the uniaxial anisotropy in $\vvi$ and $\cri$.
Without including SOC on I site, $\cri$ has a negligibly small easy-plane anisotropy from $K_\text{Cr-Cr}$ while $\vvi$ exhibits an even larger easy-plane MAE from $K_\text{V-V}$.
A small positive Cr-I contribution $K_\text{Cr-I}$ in $\cri$; a 9\% decrease in MAE is observed when one removes SOC from Cr in $\cri$.
Our DFT results for $\cri$ are consistent with previous calculations~\cite{lado20172m,Xu2018cm}.
Remarkably, the much smaller MAE in $\vvi$ than in $\cri$ results mostly from a large negative inter-sublattice contribution $K_{\text{V}-\text{I}}$; which had also been found in $L1_0$ FeNi and FePd~\cite{ke2019prb,blanco-rey2019njp}.
Turning off SOC on V increases the MAE in $\vvi$ by 335\%.

SOPT can be very useful to describe and understand MAE in various systems~\cite{ke2015prb,ke2016prbA}.
\rFig{scale} also shows the quadratic fits of MAE values, on the basis of Eq.~\eqref{eq:secorpergen}, using data calculated for $\lambda\in[0, 0.4]$, and corresponding extrapolation to $\lambda=1.5$.
The deviation of the fit from the calculated data indicates the validity of the SOPT treatment of the SOC and MAE.
For small perturbations (small $3d$ SOC comparing to bandwidth) one expects a perfect quadratic dependence of MAE on the scaling factors with the SOPT, as shown in the plots  in \rfig{scale} for the scaling scheme that $\lambda_{\text{V/Cr}}$ is varied while $\lambda_{\text{I}}=0$.
However, the large SOC constant of I-$p$ orbitals causes deviations of MAE from SOPT in $\cri$ and $\vvi$, especially the latter.
Comparing the fitted $K(\lambda_{\text{V}} = \lambda_{\text{I}})$ with the calculated ones, $\vvi$ shows a dramatically larger deviation of 42\% than 9\% in $\cri$.
This large deviation leads to the conclusion that SOPT describes MA effects better in $\cri$ than in $\vvi$.
To confirm this conclusion, we have also calculated MAE of $\cri$ and $\vvi$ using force theorem and SOPT with full-potential linear augmented plane wave (FP-LAPW) method~\cite{wien2k}.
The calculations produced about 75\% (18\%) different MAE values between the two different methods for $\vvi$ ($\cri$).
Note that $\vvi$ has a much smaller bandgap than $\cri$, which likely makes SOPT less valid in describing the MAE in $\vvi$.

\section{CONCLUSIONS}
In summary, we have theoretically studied the electronic and magnetic properties of the two-dimensional (2D) van der Waals magnet VI$_3$ in various forms.
The interlayer exchange couplings show a stronger dependence on the dimensionality of materials than the intralayer couplings.
Moreover, $\vvi$ also shows a strong stacking-dependent interlayer coupling, similar to $\cri$,  rendering the bilayer ground state interlayer ordering readily switchable between the FM and AFM configurations upon external stimuli, such as strain and electric field.
Considering the relative positions of and the hybridizations between cation-$3d$ and anion-$p$ states can be essential to determine the superexchange coupling in $\vvi$, future studies going beyond DFT+$U$ and including non-local exchange correlations will be useful to investigate the magnetic interactions in $\vvi$.
Within SRSWT, the calculated $\tc$ in bulk agrees well with experiments while $\tc$ in the monolayer and bilayer are quenched down to 14 and 24 K, respectively.
The sublattice-resolved MAE calculated in the monolayer $\vvi$ reveals that a strong V-I inter-ion easy-plane contribution to MAE lowers the overall easy-axis MAE in $\vvi$.
Finally, we demonstrate that the second-order perturbation treatment of the MAE is less valid in $\vvi$ than in $\cri$, due to the combination of large SOC on I sublattice and a smaller bandgap in $\vvi$.

\section*{ACKNOWLEDGMENTS}
The authors thank B.~Harmon, P.~Ong and G.~Miller for valuable discussions.
This work was supported by the U.S.~Department of Energy, Office of Science, Office of Basic Energy Sciences, Materials Sciences and Engineering Division, and Early Career Research Program.
Ames Laboratory is operated for the U.S.~Department of Energy by Iowa State University under Contract No.~DE-AC02-07CH11358.

\appendix

\section{Intralayer coupling}
For monolayer, the energies (per magnetic atom) of FM, N\'eel-, zigzag-, and stripy-AF configurations, $E_\text{FM}$, $E_\text{N}$, $E_\text{Z}$, and $E_\text{S}$, respectively, are mapped onto \req{eq:Hamiltonian_e} as follows:
\begin{equation}
\label{mapping}
    E_\mu = E_0 + \sum_{k=1}^3C^\mu_kJ_k, \quad \mu = \text{FM}, \, \text{N}, \, \text{Z}, \, \text{S},
\end{equation}
where $E_0$ is the non-magnetic energy, and the coefficients $C^\mu_k$ are listed in \rtbl{tbl:spin_coefficients}.

\begin{table}[htbp]
\caption{Coefficients of mapping of different spin configurations onto the spin Hamiltonian defined as $H=-\sum J_{ij}\hat{e}_i\cdot\hat{e}_j$.
The right columns represent the difference, $\Delta C^\mu =C^\mu-C^{\text{FM}}$.}
\label{tbl:spin_coefficients}
\bgroup
\def\arraystretch{1.2}%
\begin{tabular*}{\linewidth}{l @{\extracolsep{\fill}} rrrrcrrr}
  \hline \hline
\multirow{2}{*}{Configurations}  & \multicolumn{3}{c}{$C$}       &  & \multicolumn{3}{c}{$\Delta C$}   \\ \cline{2-4}  \cline{6-8}
                & $C^\mu_1$ & $C^\mu_2$ & $C^\mu_3$ &  & $\Delta C^\mu_1$ & $\Delta C^\mu_2$ & $\Delta C^\mu_3$  \\ \hline
FM              &    -3 &    -6 &    -3 &  &     0 &     0 &     0  \\
N\'eel          &     3 &    -6 &     3 &  &     6 &     0 &     6  \\
Zigzag          &    -1 &     2 &     3 &  &     2 &     8 &     6  \\
Stripy          &     1 &     2 &    -3 &  &     4 &     8 &     0  \\ \hline
\end{tabular*}\\
\egroup
\end{table}

\bibliography{References}

\end{document}